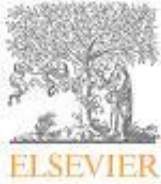

# A high-level synthesis approach for precisely-timed, energy-efficient embedded systems

Yuchao Liao*, Tosiron Adegbija, and Roman Lysecky

*Electrical and Computer Engineering, University of Arizona, Tucson, 85721, USA*

**Abstract**

Embedded systems continue to rapidly proliferate in diverse fields, including medical devices, autonomous vehicles, and more generally, the Internet of Things (IoT). Many embedded systems require application-specific hardware components to meet precise timing requirements within limited resource (area and energy) constraints. High-level synthesis (HLS) is an increasingly popular approach for improving the productivity of designing hardware and reducing the time/cost by using high-level languages to specify computational functionality and automatically generate hardware implementations. However, current HLS methods provide limited or no support to incorporate or utilize precise timing specifications within the synthesis and optimization process. In this paper, we present a *hybrid high-level synthesis (H-HLS)* framework that integrates state-based high-level synthesis (SB-HLS) with performance-driven high-level synthesis (PD-HLS) methods to enable the design and optimization of application-specific embedded systems in which timing information is explicitly and precisely defined in state-based system models. We demonstrate the results achieved by this H-HLS approach using case studies including a wearable pregnancy monitoring device, an ECG-based biometric authentication system, and a synthetic system, and compare the design space exploration results using two PD-HLS tools to show how H-HLS can provide low energy and area under timing constraints.

*Keywords:* High-level synthesis; periodic state-machines; precise timing specification; energy optimization; embedded systems

## 1. Introduction

Embedded systems continue to have widespread and growing importance across very diverse applications, including medical devices, autonomous vehicles, agriculture, telecommunications, etc. Embedded systems are growing even more popular with the rapid growth of the Internet of Things (IoT). Many embedded systems require the use of application-specific hardware to meet precise timing requirements within stringent resource constraints, including area and energy. The design and optimization of these application-specific embedded systems requires careful consideration of precise timing constraints, reliability, security, and energy consumption. For life-critical embedded systems, such as wearable or implantable medical devices, the need to achieve precise real-time sensing, computation, and communication constraints while optimizing energy consumption is a critical challenge

* Corresponding author. E-mail address: yuchaoliao@arizona.edu (Y. Liao), tosiron@arizona.edu (T. Adegbija), rlysecky@arizona.edu (R. Lysecky).

[10]. Furthermore, development costs, regulatory compliance, and time-to-market pressures necessitate rapid design and optimization methods.

High-level synthesis (HLS) is a popular approach to the design and synthesis of these application-specific hardware implementations. HLS allows designers to automatically develop hardware implementations from a high-level language (e.g., C, C++) instead of implementing in low-level hardware description languages (HDLs) (e.g., Verilog, VHDL). HLS enables a system to be designed at a higher level of abstraction, which can significantly increase productivity while reducing development times/costs. Studies have shown that designer effort can be reduced by up to 6X with HLS-style development, compared to using Verilog/VHDL to design hardware [22].

However, existing HLS approaches have a few disadvantages and limitations that this work aims to address. First, specifications in high-level languages rely on the implied order of operations that requires timing to be inferred from the structure or manually added through tool-specific annotations [22][23][35]. To effectively use these annotations, designers must have extensive understanding of hardware design and synthesis algorithms [1]. Second, HLS approaches primarily target synthesizing individual computations to high-performance hardware implementations, where the output of the HLS tools is a set of individual hardware components. Thus, system designers must integrate the synthesized hardware components to construct the overall system before considering system-level optimization. Third, and most importantly, existing HLS approaches provide limited or no support to incorporate or utilize precise timing specifications within the synthesis and optimization process [35]. As such, even though current HLS approaches substantially ease the register transfer-level (RTL) design process, they are limited in how much control the designer has on timing accuracy, the ability to fine-tune the output design, and the ability to optimize the design with respect to system-level timing constraints.

In this paper, we present **Hybrid HLS (H-HLS)**, a new holistic HLS methodology for designing and optimizing application-specific systems where precise timing behaviors are inherent and critical to the correct system operation. H-HLS augments the traditional HLS approach, which we call *performance-driven HLS (PD-HLS)*, with a novel *state-based HLS (SB-HLS)* method that enables the design and optimization of application-specific embedded systems in which timing information is explicitly and precisely defined. PD-HLS enables the synthesis of complex multicycle computations with Pareto-optimal performance, power, and area tradeoffs, while the SB-HLS supports the synthesis of the timing behaviors via precise timing specifications to automatically extract *component-level* and *system-level* synthesis constraints. With the explicit and precise timing in component and system models, H-HLS allows application-specific hardware that executes only as fast as needed with minimal energy. We demonstrate the benefits of the H-HLS approach using two case studies—a wearable pregnancy monitoring (WPM) device and an ECG-based biometric authentication (EBA) system—and a synthetic system to show that low energy consumption and low area was achieved from H-HLS under system-level time constraints. We evaluate our H-HLS methodology using two different PD-HLS tools: Microchip's LegUp HLS tool, to represent the state-of-the-art, and a custom-built latency-constrained force-directed scheduling (LC-FDS) HLS tool that implements the force-directed scheduling algorithm [28] and supports explicit precise timing specification.

Our main contributions are summarized as follows:

- We present a novel Hybrid HLS (H-HLS) methodology that combines a novel state-based HLS (SB-HLS) and performance-driven HLS (PD-HLS).
- H-HLS incorporates a notion of explicit and precise timing information using novel periodic state machine (PSM) formalisms.
- We propose a new SB-HLS approach that analyzes the PSM models to identify precise timing information, identify single-cycle computations that can be directly synthesized, and identify multi-cycle computations to be synthesized using PD-HLS.
- We have developed a custom-built HLS tool (called LC-FDS) that enables explicit and precise timing information. We evaluate our approach in the context of the state-of-the-art using

Microchip's LegUp HLS tool, which does not support precise timing specification, to demonstrate H-HLS's effectiveness as a holistic approach for designing precisely-timed, energy-efficient application-specific systems.

- To evaluate the proposed H-HLS approach, we implemented two detailed case studies of a wearable pregnancy monitoring (WPM) device and an ECG-based biometric authentication (EBA) system. We additionally implemented a synthetic system with greater complexity to evaluate the scalability of the proposed approach. Results show that compared to an unoptimized design, the H-HLS approach yields average energy reductions of 92.9%, and up to 94.6%. Compared to LegUp, our custom-built tool (LC-FDS) reduces the energy and area by an average of 41.9% and 59.6%, and up to 48.2% and 76.7%, respectively, demonstrating the benefits of the proposed approach.

## 2. Background and related work

High-level synthesis (HLS) [7][9][16] provides an abstraction of programming effort above the register-transfer level (RTL). HLS techniques and tools are growing in popularity because they substantially increase designer efficiency. HLS allows hardware to be designed using high-level software programming languages like C/C++, rather than high-expertise low-level hardware description languages like VHDL/Verilog [16][4]. However, HLS has limitations that necessitate new abstractions for designing emerging application-specific embedded systems. Studies [22][17][34] show the state-of-the-art HLS tools still lag manual RTL design in the quality of results produced. Even though HLS tools substantially ease the RTL design process, they are limited in how much control the designer has on timing accuracy and the designer's ability to fine-tune the design output. HLS tools require that the software be written in specific ways to exploit the inherent fine-gained concurrency that is available in hardware. Even though the tools may still map sequential software to hardware, the resultant hardware realization is typically sub-optimal with respect to performance and area [15]. Furthermore,

HLS tools use software, wherein timing is only implied in the sequence of instructions. However, embedded systems typically feature synchronous hardware that have specific timing constraints, with clock-level synchronization. Even though current HLS tools can effectively schedule and pipeline sequences of operations, they are less efficient with complex synchronizations [2]. The approach presented herein provides an important foundation for enabling HLS methods in which timing constraints can be precisely specified in the design of application-specific, low-power embedded systems.

Our work follows a *model-based design* approach for enabling precise timing specifications. Model-based design [20] plays a pivotal role in the tractability of the design process for complex systems. Models provide useful abstractions for complex and largely stochastic processes by formalizing essential properties and component interactions. For embedded systems, model-driven development faces additional challenges due to component heterogeneity, physical process concurrency, and dynamics in software, networks, and the physical world [13][36]. The need for accurate embedded systems models has spurred several research efforts in this domain [24][27]. A popular research strand focuses on developing temporal semantics for bridging the gap between discrete-time computational processes and continuous-time physical dynamics (e.g., [12][30]). Other approaches to embedded systems modeling focus on semantic definitions: formalisms, languages, and tools to help system designers develop systems that meet operational design goals [30][5]. As a useful simplification, an embedded system can be viewed as a set of heterogeneous interacting components (separation of concerns), whose behavior can be individually modeled (e.g., agent-based modeling [25][26], component-based development [6][15][18]). However, these models are not directly compatible with existing HLS tools.

A notable effort in the direction of precisely-timed systems is the idea of precision-timed (PRET) machines [14]. PRET machines were proposed as necessary for real-time systems, since traditional microarchitectures focus on superior average-case performance, potentially at the expense of predictability and repeatability of timing properties.

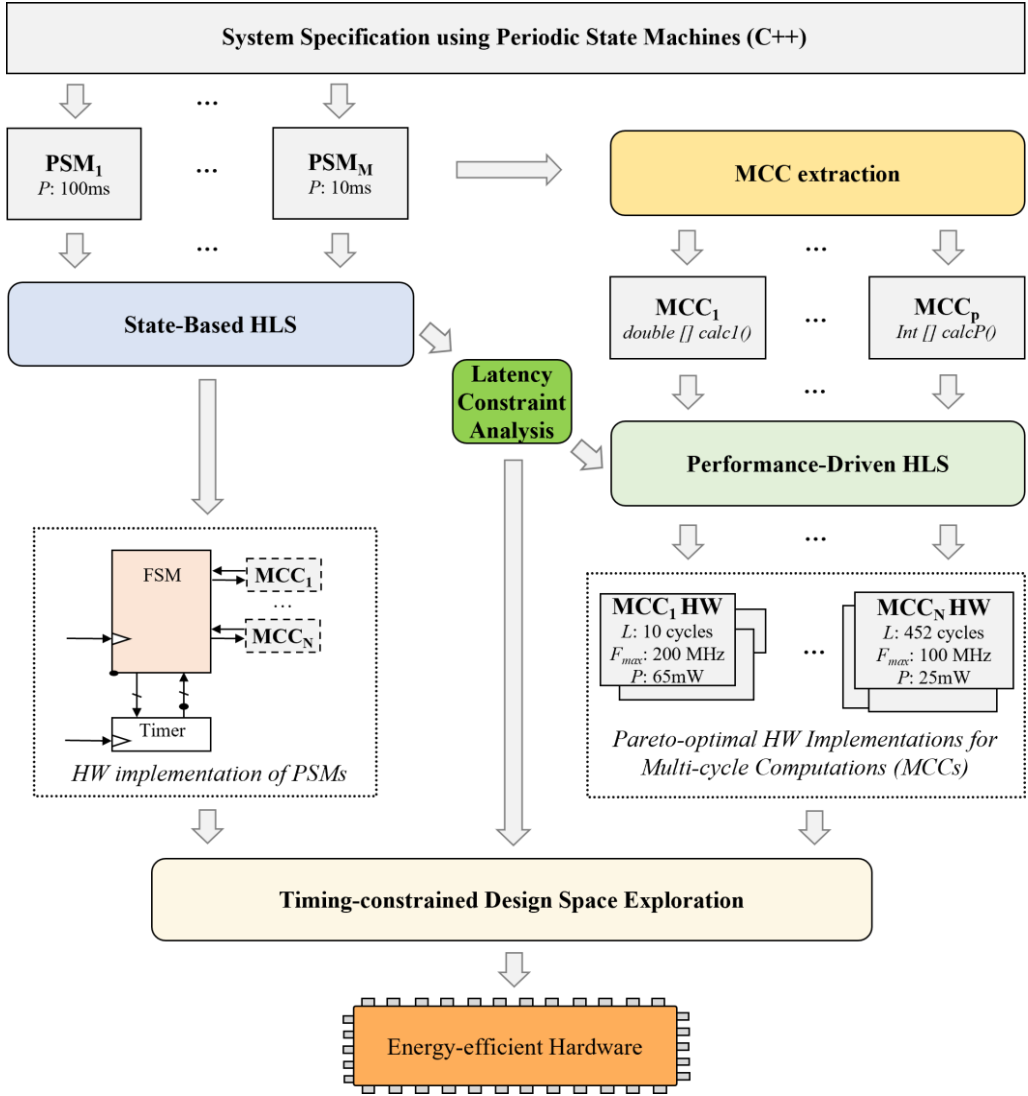

Fig. 1. Hybrid high-level synthesis (H-HLS) methodology integrating SB-HLS methods for periodic state machines, PD-HLS methods for synthesis of multicycle computations, system-level integration of SB-HLS and PD-HLS outputs, and timing-constrained design space exploration for application-specific embedded systems.

As such, PRET machines seek to deliver repeatable and controllable timing down to the precision of a clock cycle. PRET machines are microarchitectures that are capable of interrupt-driven I/O that does not disrupt timing-critical tasks. Multiple processors have been developed based on the premise of PRET machines, especially within the context of real-time and mixed-criticality systems. For example, FlexPRET [37] is a processor designed for mixed-criticality systems that uses fine-grained multithreading with flexible scheduling and timing instructions to allow executing tasks to make tradeoffs between hardware-based isolation and efficient processor utilization, depending on the tasks' criticality. Similarly, T-CREST [33] is a time-predictable multicore architecture that was designed to be optimized for worst-case execution time (WCET) performance, while easing the strain of

Table 1

State-of-the-art HLS methodologies and their key features in comparison to H-HLS

| Methodology | Precise timing specification | Precise time hardware implementation | Diversity/hardware alternatives | State based model | Frequency scaling | Design space exploration |
|---|---|---|---|---|---|---|
| [34] | | | ✓ | | | ✓ |
| [1][35] | | | ✓ | ✓ | | |
| [4][5][6][11] [15][17][18] [23][25][26] | | | ✓ | | | |
| [7] | | | ✓ | ✓ | | ✓ |
| [12][30] | ✓ | | | ✓ | | |
| [14][33][37] | ✓ | | | | | |
| H-HLS (Our approach) | ✓ | ✓ | ✓ | ✓ | ✓ | ✓ |

static WCET analysis. However, despite the importance of these research efforts, they are not suitable for the stringently resource-constrained embedded systems targeted in our work. These research efforts would generate hardware implementations that are overprovisioned for applications with similar functional characteristics and design constraints to the target devices in our work. The work presented herein is not intended as a replacement to this existent body of work, but rather as an orthogonal contribution to a different and important class of specialized applications in which high-level synthesis is used to create precisely-timed, energy-efficient, application-specific hardware for resource-constrained embedded systems.

Our H-HLS methodology incorporates the features highlighted in Table 1 to enable the holistic design of highly efficient application-specific embedded systems. These features are highlighted within the context of several related state-of-the-art HLS methodologies to illustrate the benefits of H-HLS in comparison to these methodologies. These features are briefly described as follows:

- *Precise timing specification*: The ability to precisely define component- and system-level timing constraints, achieved in this work via novel periodic state machine (PSM) formalisms.
- *Precise time hardware implementation*: The ability to implement a final system-level hardware that meets the precise time behaviors defined in the PSMs.
- *Diversity/hardware alternatives*: The methodology's ability to provide alternative hardware implementations that can be explored to best satisfy system requirements.

- *State-based model*: The use of the state-based modeling abstraction to provide a close representation of the target algorithmic entity and provide a convenient framework for designing, analyzing, verifying, and validating the system.
- *Frequency scaling*: The use of the minimum operating frequency for each multi-cycle computation, such that the power consumption is minimized while meeting the precise timing requirements of each PSM.
- *Design space exploration*: Exploration of the design space of hardware alternatives to determine the hardware combinations that best satisfy system requirements.

## 3. Hybrid high-level synthesis methodology

Fig. 1 provides an overview of the H-HLS methodology. The H-HLS methodology is distinguished from existing approaches by providing a formal design, synthesis, and optimization framework for precisely-timed, application-specific embedded systems. Our H-HLS approach meets precise-timing, energy-efficient, and application-specific hardware requirements by combining SB-HLS and PD-HLS methods. The H-HLS methodology uses a synthesizable periodic state machine (PSM) model implemented in C++ to specify the behavior of individual system components. The PSM model enables the explicit specification of precise timing behaviors and use of event-based communication/synchronization between PSMs. For each PSM component model, SB-HLS identifies states, timing behaviors, input/output events, single-cycle operations, and inter-PSM synchronization to synthesize a register-transfer level (RTL) implementation in a hardware description language (e.g., Verilog/VHDL). During this synthesis process, simple computations (e.g., multiplication) that can be executed in single clock cycle are directly implemented as RTL operations. For more complex multi-cycle computations (MCCs), PD-HLS is used to synthesize multiple hardware accelerators with Pareto-optimal performance, power, and area tradeoffs. Given the resulting RTL implementations for PSMs and Pareto-optimal hardware alternatives for MCCs therein, timing-constrained design space exploration is used to find energy-efficient system implementations that satisfy the precise timing constraints. Importantly, H-HLS utilizes the PSMs' precise timing specifications to automatically extract component-level and system-level synthesis constraints used during design space exploration.

### 3.1. Periodic State Machine (PSM)

Periodic state machine (PSM) [21] modeling formalism is foundational for the H-HLS methodology. Traditional PSM enables the specification of time triggered execution of state machines defined by a fixed execution period. We have augmented the PSM modeling approach to also support explicit and precise timing information, explicit event-based specifications of data communication and synchronization between PSMs, and specification of both simple single-cycle computations (e.g., addition, comparison) and more complex multicycle computations (e.g., matrix multiplication, median filter).

The PSM supports system-level specification through hierarchical specifications of component and system models. Component models specify the behavior of individual components. System models specify the instances and connections between component models and consist of ports, component model instances, and port connections. The input and output ports define a model's external interface, through which component and system models can be connected to other models.

A PSM component model consists of a set of finite states representing the system behavior. Inputs define the events the component model will react to, and outputs define the events that can be generated by the model. Inputs and outputs can also be associated with data values that are transmitted by the events. An *external transition* changes the state in response to an input event. Each state can have an external transition for each input to determine the next state of the model using an *import* statement. Conversely, *internal transitions* are generated internally within the model based on either conditional expressions or timing specifications. While a PSM uses a fixed period to define the fastest rate at which the PSM needs to execute, internal timing specifications can be used to further specify precise execution behaviors.

A *timing specification* can describe the amount of time spent in a specific state without receiving an external event. Timing specifications can also have two special cases: an *infinite time specification* and a *delta time specification*. An infinite time specification (e.g., $ts(\infty)$) indicates that the state will not have an internal transition and can only transit to other states based on external events. A delta time specification (e.g., $ts(\Delta)$) indicates that no time is spent in the current state, which implies that any internal transition or output events are immediately generated.

Communication of data between component models in PSMs is handled by pairing data transfers with events—referred to as data events. For example, when an input event with data is received, the external transition handling the event can read the data as though it were an internal variable.

Finally, the PSM's actions include a sequence of operations consisting of either output events, simple single-cycle computations, or multicycle computations. Output events are specified using *notify* statements for non-data events and *export* statements for data events. Simple single-cycle computations are those operations that can be implemented in hardware within a single clock cycle, such as a simple multiplication. In contrast, multicycle computations (MCCs) require multiple clocks to execute, such as computing the median value within a sample of data measurements. This distinction is exploited in the H-HLS approach by using the SB-HLS to handle single-cycle computations and using the PD-HLS to handle MCCs. For simplicity, we assume that MCCs are

**MaternalHRMonitor PSM**
*P*: 100 ms; *Inputs*: Start; *Outputs*: sampleMHR, avgMHR, Done, bcAlert

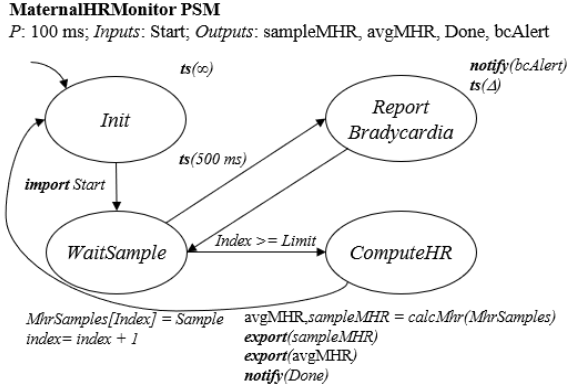

Fig. 2. Illustrative PSM for a simplified maternal heart rate monitoring component within a wearable pregnancy monitoring device.

implemented as functions and single-cycle computations are implemented as assignment statements using built-in operators.

Fig. 2 illustrates a simplified component-level PSM for a maternal heart rate monitor component within a wearable pregnancy monitor (WPM) device, which is described further in Section 4. The maternal heart rate (MHR) PSM periodically measures, logs, and analyzes the maternal heart rate to detect conditions such as bradycardia (abnormally slow heart rate) and tachycardia (abnormally fast heart rate). The *start* input is an event that indicates the start of a new maternal heart rate measurement without data transfer in this event. The state transition from *Init* to *WaitSample* uses an *import Start* event specification indicating that the *Init* state should transition to *WaitSample* when the *Start* input event is received. The state transition from *WaitSample* to *ReportBradycardia* illustrates the use of a time specification, where *ts*(500 ms) indicates that if no sample is received within 500 ms, the PSM should transition to *ReportBradycardia* to trigger an alert that the heart rate is too slow. A *ReportTachycardia* state is omitted from the figure, but functions similarly to *ReportBradycardia* when samples are received more frequently than a threshold.

### 3.2. State-based HLS (SB-HLS)

Given the PSM model, the SB-HLS methodology exploits the precise timing information captured within the PSM models to create synthesizable HDL-based implementations. SB-HLS extracts the components in the PSM model, some of which have direct paralles in hardware design (e.g., the concept of states), for direct synthesis in RTL. During synthesis, the PSM description of the states, transitions between states, and single-cycle computations can be directly used to generate a *finite state machine* (FSM). Other PSM capabilities, such as time specifications and the communication of events between models require integration of specialized hardware components, including input/output event handshake and timer components.

During the synthesis of component models, the SB-HLS uses the duration of time specifications to determine how it is implemented in hardware. Time in PSM models falls into one of three categories: infinite time specifications, finite constant time specifications, and delta time specifications. Infinite time specifications result in no action in either hardware component. Finite constant time specifications can be achieved by using a timer component that is configured from the FSM to count from zero to the specified time. Delta time specifications are realized as a single clock cycle.

Within PSM specification, time is explicitly defined using real-time values (e.g., milliseconds). However, within the synthesized hardware implementations, time is represented in clock cycles. Thus, the SB-HLS process must convert the explicitly defined time to clock cycles. As different instances of component models may execute using different clock frequencies, the SB-HLS utilizes generics/parameters within the HDL description for finite time specifications. These generics can be configured for each PSM component model instance within the synthesis of system models. This allows different instances of the same PSM component model to operate at different frequencies while achieving the same timing.

### 3.3. Performance-driven HLS (PD-HLS)

The PD-HLS phase of our H-HLS methodology involves using existing HLS tools to synthesize and generate multiple alternative hardware implementations for each MCC defined within a PSM. PD-HLS uses the explicit timing specifications within PSMs to automatically define fine-grained latency constraints to create synthesized hardware implementations and explore their performance/energy tradeoffs. We consider two different HLS tools for the PD-HLS: Microchip's LegUp [23] HLS tool (now called SmartHLS) and a custom-built latency-constrained force-directed scheduling (LC-FDS) based HLS tool.

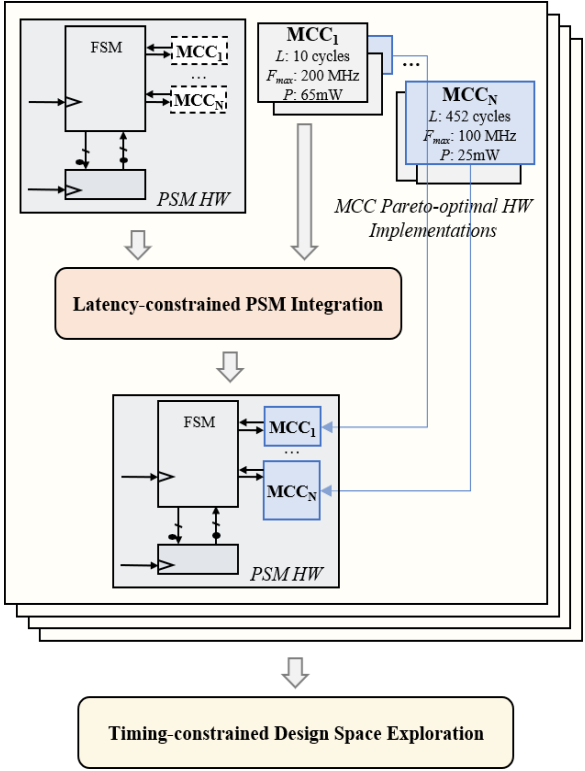

Fig. 3. Timing-constrained design space exploration determines Pareto-optimal configurations by exploring MCC hardware alternatives and frequency scaling using time constraints defined within the PSM models.

The LegUp HLS tool is primarily focused on maximizing performance given resource constraints, and not on minimizing energy consumption given latency constraints. Latency and energy exploration can still be achieved but must be done indirectly via resource constraints. LegUp uses standard techniques for high-level synthesis, consisting of compiler-optimization, scheduling, and resource binding. During the scheduling stage, a designer can specify resource constraints, which are typically used to reduce area. Tighter resource constraints (e.g., fewer adders allowed) may result in longer execution times (i.e., a higher number of clock cycles required to perform the computations). By exploring different resource constraints for each MCC being synthesized, designers can indirectly create hardware implementations with different latency, area, and energy tradeoffs.

The LC-FDS HLS tool utilizes a latency-constrained forced-directed scheduler [28] that, given a latency constraint, determines an execution schedule with minimum required resources. Using different latency constraints, multiple MCC implementations can be determined with tradeoffs in latency and area. On one extreme, the H-HLS approach can determine the minimum achievable latency constraint, which would result in the best performing hardware implementation with the largest area requirement. On the other extreme, the longest latency constraint that does not result in idle cycles in the execution schedule can be determined. Such a constraint would result in the lowest performance implementation with the minimum area requirement. Additional latency constraints between these two extremes can be explored. Using a range of latency constraints, LC-FDS can create multiple alternative hardware implementations for each MCC.

### 3.4. Timing-constrained design space exploration

Fig. 3 presents an overview of timing-constrained design space exploration (DSE) for determining Pareto-optimal system configurations. During the synthesis of the hardware for each PSM, the SB-HLS generates a finite-state machine implementation for the PSMs with placeholders for the MCCs. The PD-HLS tool generates multiple alternative hardware implementations for each MCC, which may have different execution latencies and maximum operating frequencies. The latency-constrained PSM integration block integrates the hardware for the synthesized MCC within the hardware implementations for the PSM. The latency-constrained PSM integration task integrates the PSM's HW implementation with the selected hardware alternative for each required MCC to create the final complete HW implementation for the PSM. This task includes using the explicit time constraints captured in the PSM model to scale the operating frequency for the MCCs in each PSM. The goal of frequency scaling is to determine the operating frequency for the MCCs such that the PSM's precise timing behavior is achieved, while minimizing each PSM's energy consumption. Furthermore, this task also includes configuring the required communication and synchronization of the FSM with the MCC hardware.

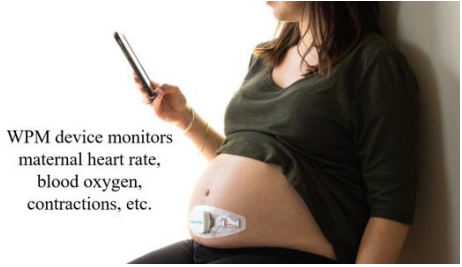

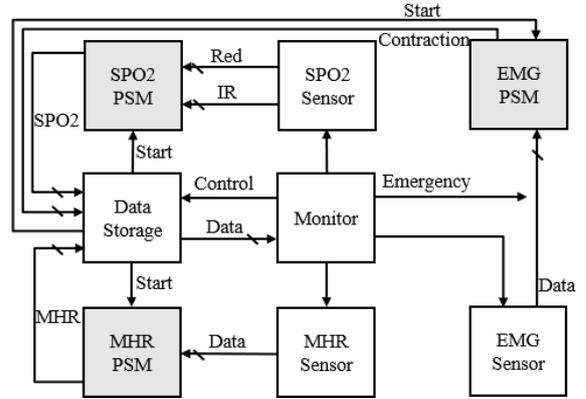

Fig. 4. Wearable pregnancy monitoring (WPM) device attached to a mother's abdomen monitors vitals of mother. WPMs require energy efficiency and high-performance necessitating application-specific hardware. Photo: Bloomlife, 2021.

At the system level, the resulting configurations will have tradeoffs in terms of total area requirements and energy consumption. Different design space exploration methods can be used, depending on the complexity of the design space. For a tractable design space (e.g., with hundreds of solutions), exhaustive search can be used to identify the best system configuration. However, for larger and more complex systems with millions of possible solutions, more complex methods, like genetic algorithms or machine learning models that are amenable to multi-objective optimization problems can be used.

## 4. Case studies

This section describes two case studies and one synthetic system used to illustrate the H-HLS approach. The case studies include two real-world systems: a wearable pregnancy monitoring (WPM) device and an ECG-based biometric authentication (EBA) system. The synthetic system is a generic system that abstracts out the low-level application-specific implementation details of the aforementioned systems and increases the system complexity in terms of PSM states, number of MCCs, and number of MCC alternatives. To demonstrate the benefits of the H-HLS approach, we opted to use realistic case studies, rather than multiple benchmarks, as the case studies enable us to evaluate the practical system-level implications of the H-HLS approach that would not be captured by using multiple different benchmarks. While the same analysis was performed for all our experiments, for brevity, the WPM device is described in detail, while the EBA and synthetic systems are described in summary to provide additional context for our work.

Fig. 5. System-level PSM overview of a wearable pregnancy monitoring device used to demonstrate the H-HLS modeling, synthesis, and optimization.

### 4.1. Wearable pregnancy monitoring (WPM) device

WPM devices are intended to improve access to maternal healthcare in rural and underserved communities with limited healthcare access. Fig. 4 illustrates a WPM device that attaches to a mother's abdomen to detect and monitor vital signals during pregnancy. Some of the design requirements for this WPM device include: 1) small physical size; 2) high performance to execute computationally complex algorithms; 3) precisely defined timing constraints; and 4) very low energy consumption to ensure long operation. We used the H-HLS approach to design a WPM device that measures maternal heart rate (MHR) [29], blood oxygen saturation (SPO2) [32], and abdomen contraction electromyography (EMG) [3]. Apart from the system-level evaluation, these three kernels also enable us to evaluate the H-HLS approach across different workloads with different computational characteristics.

Fig. 5 presents the system-level overview of the WPM device. The system consists of seven

Table 2

Components and main functions in the wearable pregnancy monitoring (WPM) system, ECG-based biometric authentication (EBA) system and synthetic system.

| System | Component | Main Functions | MCC Alternatives (Sum in PSM) | Number of States (Non-unrolled, minimum latency) | Number of Loops | Number of Iterations (Outer+Inner) |
|---|---|---|---|---|---|---|
| Wearable pregnancy monitor (WPM) device | MHR | Filter, Peak detection, MHR average | 2 | 68 | 2 | 750 |
| | SPO2 | Filter, SPO2 calculation | 8 | 58 | 1 | 1000 |
| | EMG | Filter, Irregular peak detection | 5 | 216 | 7 | 105622 |
| ECG-based Authentication (EBA) system | Signal filtering | Filter the input signal | 2 | 78 | 1 | 1000 |
| | Segmentation | Detects R-Peak, Perform segmentation | 9 | 280 | 10 | 60198 |
| | Feature extraction | Detects fiducial points | 6 | 125 | 5 | 145 |
| Synthetic System | PSM1 | 3 MCCs | 8 | 390 | 36 | 92493 |
| | PSM2 | 3 MCCs | 6 | 1179 | 75 | 385062 |
| | PSM3 | 2 MCCs | 5 | 560 | 32 | 160856 |
| | PSM4 | 3 MCCs | 8 | 417 | 18 | 322260 |

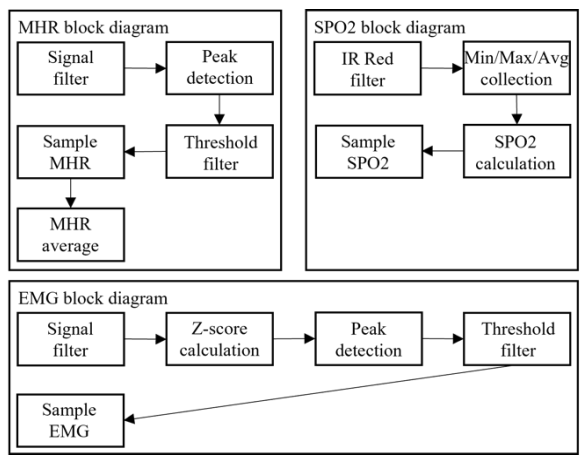

Fig. 6. Functional block diagrams for a wearable pregnancy monitoring device's computation, MHR, SPO2, and EMG.

components, and input and output signals between the components. The system has MHR, SPO2, and EMG sensors, which frequently sample signals from mother's abdomen. The corresponding PSM component for each sensor is used to take measurements at the required intervals and compute the current MHR, SPO2, and EMG values. The data storage component stores the sensor data and the communication signals between PSM components and monitor component. The monitor component analyzes all sensor data and sends alerts and notification, when necessary, if a signal deviates from normal values.

Table 2 presents the main computations in each case study. The table also highlights the number of MCCs generated for each PSM, the number of states within each PSM, the number of loops in each computation, and the number of iterations including both the outer and inner loops. Fig. 6 illustrates the function block diagrams for the computations within the MHR, SPO2, and EMG PSMs. The MHR computation analyzes and filters the input signals until it finds a valid heart rate. MHR uses thresholds to find each valid heartbeat (a peak value) and uses the time between heartbeats to calculate a valid heart rate. MHR computation outputs the average heart rate and sample heart rate to the MHR PSM. SPO2 computation analyzes and filters IR and RED signals to find important values like the maximum and average of the signals. SPO2 computation uses those values to calculate and output the blood oxygen level to the SPO2 PSM. Compared with the MHR computation, SPO2 allows more parallel operations, thereby enabling the exploration of a larger design space. The EMG computation analyzes and filters the input signals to find irregular peaks of the signals to detect contractions. To find the irregular peaks, EMG computation calculates the Z-score each time there is a new input signal. Z-score calculation takes the most of time in the EMG computation. EMG computation features more parallel operations and higher complexity than MHR and SPO2 computations.

For the design space exploration of the hardware alternatives generated using our approach in this case study, we exhaustively explored the entire design space to evaluate the Pareto-optimal trade-offs between area and energy consumption. Specifically, we exhaustively explored the system-level implementations for all possible combinations of the hardware alternatives for the MCCs within the MHR, SPO2, and EMG PSMs. For each system-level implementation, we determined the minimum operation frequency for each MCC that still meets the precise timing requirements of each PSM. Finally, we manually determined which implementations were Pareto-optimal in terms of area and energy consumption. Note that the tractability of the design space in this case study allowed us to perform exhaustive evaluations and manual analysis to determine efficient system configurations.

### 4.2. ECG-based biometric authentication (EBA) system

ECG-based biometric authentication (EBA) is a popular approach for biometric authentication in consumer devices. EBA systems offer several attractive qualities, including the individualized, ubiquitous, and easily identifiable nature of ECG signals. The EBA system is comprised of four main computations: *filtering, segmentation, feature extraction,* and *matching.* As depicted in Table 2, our EBA system is implemented using three PSMs for filtering, segmentation, and feature extraction. We assume that the matching computation is executed solely on the CPU. The filtering PSM processes the gathered ECG signals to remove noise and enhance the quality of the biometric traits. The segmentation PSM detects R-peaks in the filtered signal and splits the ECG signal into its unique component waveforms to reduce redundancy in the signals. The feature extraction PSM detects fiducial points in the signal and extracts these features to enable the system to distinguish between different users. Compared to WPM system, the computations in the EBA system feature more nested loops and if-else statements, as depicted in Table 2. For more low-level details of the EBA algorithm, we direct the reader to [8].

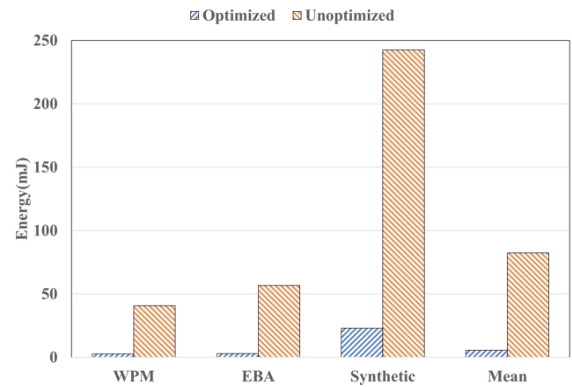

Fig. 7. Optimized vs. Unoptimized system-level energy consumption for a wearable pregnancy monitoring (WPM) device, an ECG-based biometric authentication (EBA) system, a synthetic system, and a geometric mean by using LC-FDS tool.

### 4.3. Synthetic system

To further analyze the H-HLS approach, we generated a synthetic system to model the structural characteristics of a generic system that is abstracted from the application-specific implementation details of the WPM and EBA systems. This synthetic system is especially instructive for our analysis of the timing-constrained design space exploration for determining Pareto-optimal system configurations. The synthetic system is comprised of four PSMs with three, three, two, and three multi-cycle computations in each PSM, as shown in Table 2. The computational characteristics of the synthetic system is randomly generated based on a wide range of characteristics derived from real-world systems.

## 5. Case study experimental results

To evaluate the PSM formalism and H-HLS approach, we synthesized multiple implementations of each MCC within the WPM's maternal heart rate (MHR) PSM, blood oxygen saturation (SPO2) PSM,

Table 3

PD-HLS synthesis results for MCCs within the wearable pregnancy monitoring (WPM) system and the ECG-based biometric authentication (EBA) system using the LC-FDS and LegUp HLS tools.

| System | MCC | LC-FDS | | | | | | LegUp | | | | |
| | | Constraints | | Synthesis Results | | | | Constraints | Synthesis Results | | | |
| | | Loop Unrolling | Latency (Cycles/ Iteration) | Freq. (MHz) | Total Exec. Latency (Cycles) | Area (LUT+FF) | Power (mW) | Loop Unrolling | Freq. (MHz) | Total Exec. Latency (Cycles) | Area (LUT+FF) | Power (mW) |
|---|---|---|---|---|---|---|---|---|---|---|---|---|
| WPM | MHR | 0 | 63 | 102 | 4056 | 1397 | 135 | 0 | 121 | 3406 | 4277 | 139 |
| | | 4 | 177 | 105 | 3527 | 1510 | 139 | 4 | 120 | 3928 | 5570 | 150 |
| | SPO2 | 0 | 56 | 103 | 5055 | 3046 | 147 | 0 | 119 | 9050 | 3354 | 147 |
| | | 4 | 66 | 103 | 4553 | 4270 | 149 | 4 | 124 | 6800 | 4821 | 170 |
| | | | 63 | 102 | 3803 | 4153 | 155 | | | | | |
| | | | 62 | 105 | 3553 | 4761 | 159 | | | | | |
| | EMG | 0 | 163 | 94 | 885316 | 3097 | 137 | 0 | 120 | 1907289 | 5489 | 173 |
| | | 5 | 200 | 98 | 639041 | 4742 | 156 | 5 | 122 | 1070739 | 6950 | 194 |
| | | | 198 | 97 | 629531 | 4671 | 148 | | | | | |
| | | | 196 | 96 | 620021 | 4655 | 144 | | | | | |
| EBA | Filtering | 0 | 65 | 107 | 72613 | 1048 | 115 | 0 | 106 | 152952 | 4891 | 117 |
| | | 4 | 260 | 106 | 41068 | 1393 | 127 | 4 | 106 | 80912 | 7006 | 136 |
| | Segmentation | 0 | 242 | 87 | 640849 | 5921 | 226 | 0 | 102 | 873304 | 26382 | 203 |
| | | | 239 | 92 | 601570 | 5976 | 243 | | | | | |
| | | | 238 | 86 | 562292 | 6036 | 233 | | | | | |
| | | | 236 | 87 | 483736 | 6035 | 237 | | | | | |
| | | | 228 | 90 | 483652 | 5956 | 238 | | | | | |
| | | | 225 | 90 | 483259 | 6133 | 242 | | | | | |
| | Feature extraction | 0 | 109 | 111 | 1317 | 1985 | 135 | 0 | 123 | 2112 | 10273 | 147 |
| | | 4 | 106 | 113 | 1305 | 1912 | 130 | 4 | 116 | 1531 | 6789 | 131 |
| | | | 361 | 112 | 1310 | 2326 | 136 | | | | | |
| | | | 349 | 111 | 1298 | 2290 | 137 | | | | | |

abdomen contraction (EMG) PSM, EBA's filtering PSM, segmentation PSM, and feature extraction PSM. We implemented these MCCs using both LC-FDS and LegUp. Using these tools enables us to analyze and quantify the area and energy benefits of explicit timing specifications (enabled by LC-FDS) in the context of the state-of-the-art (LegUp). We then explored the system-level area and energy tradeoffs achievable using these MCC alternatives given the explicit timing specifications within the

PSM models. All designs were synthesized using Xilinx Vivado targeting a Xilinx Artix-7 FPGA board using the xc7a100tcsg324-1 part.

Fig. 7 summarizes the overall energy results of the H-HLS approach. The figure compares the optimized systems using the H-HLS approach to the unoptimized system designed via traditional HLS [9] for the wearable pregnancy monitoring (WPM) device, ECG-based biometric authentication (EBA) system, and the synthetic system. On average, the H-HLS approach reduced the energy by 92.9%, and by up to 94.6%. In what follows, we detail the benefits and tradeoffs of the H-HLS approach with respect to performance-driven HLS and timing-constrained design space exploration.

### 5.1. Performance-Driven HLS

For both PD-HLS tools, we used loop unrolling to explore different tradeoffs in area and energy. Loop unrolling enables more parallelization of the operations and enables us to explore a larger design space for later timing-constrained design space exploration.

First, we discuss the results for the WPM device. For the MHR and SPO2 PSMs, we unrolled loops either 0 or 4 times. For the EMG PSM, we explored unrolling 0 or 5 times. The loop unrolling amounts were selected to ensure loops were unrolled as an even factor of the total loop iterations to avoid requiring pre-amble or post-amble operations.

For the LC-FDS tool, we additionally defined multiple latencies for each MCC ranging from the maximum to minimum latency constraints possible. Given the constraints defined and loop unrolling, the PD-HLS approach yielded two alternative implementations for the MHR PSM, eight for the SPO2 PSM, and five for the EMG PSM. However, for both the SPO2 and EMG PSMs, only four of the resulting hardware implementations were Pareto-optimal in terms of power consumption, area, and total execution latency. Therefore, for the SPO2 and EMG PSMs, we only considered those four hardware implementations for further exploration and analysis.

Table 3 presents the PD-HLS constraints and implementation results for the MHR, SPO2, and EMG MCCs using LegUp and LC-FDS HLS tools. For both LegUp and LC-FDS, loop unrolling mostly resulted in hardware implementations with shorter execution times, with higher power consumption, and without substantially affecting the maximum operating frequency of the resulting hardware (i.e., a marginal impact on the critical path). However, for the MHR's MCC synthesized using LegUp, unrolling yielded a design with a longer execution latency.

Given the latency constraints, the LC-FDS HLS tool reduced the latency of SPO2's execution by up to 47.8% with the tightest constraint, while also reducing the power consumption by 6.5%, compared to LegUp. With a more relaxed latency constraint, LC-FDS achieved a smaller reduction in latency of 33.0% but with a greater reduction in power consumption of 12.4%. Similarly, for the unrolled EMG MCC, LC-FDS reduced the latency and power consumption by 42.1% and 25.8%, respectively.

For SPO2 and EMG, LC-FDS reduced the area by a maximum of 13.9% and 43.6%, respectively. Even though the LC-FDS HLS tool trades off area optimization to satisfy the specified latency constraints, it was still more area-efficient than LegUp for all MCCs, while enabling the MCCs to run only as fast as needed to satisfy the latency constraints.

For the non-unrolled MHR MCC, LegUp achieved a 16.0% lower execution latency, as the latency constraint specified in LC-FDS provided some slack compared to the best latency, which LegUp prioritized. Thus, LC-FDS still satisfied the specified latency constraint and reduced the area by 67.3% compared to LegUp.

The H-HLS approach was similarly effective for the EBA system. For the filtering and feature extraction PSMs, we unrolled the loops either 0 or 4 times, as previously described, to ensure an even factor of the total loop iterations. The outer loop of the feature extraction PSM was fully unrolled four times. The loops in segmentation PSM could not be unrolled due to the presence of continue and break statements.

Table 3 summarizes the PD-HLS constraints and implementation results for the EBA system. Using the LC-FDS tool, we followed the same procedure as described for the WPM device. Given the multiple constraints and loop unrolling, the PD-HLS generated two, nine, and six alternative hardware implementations for the filtering, segmentation, and feature extraction PSMs, respectively. After finding the Pareto-optimal MCC alternatives in terms of area and energy, there were six and four hardware implementations remaining for the segmentation and feature extraction PSMs, respectively.

Given the latency constraints, the LC-FDS tool reduced the latency of execution by a maximum of 52.6% for filtering, 37.3% for segmentation, and 32.8% for feature extraction, compared to LegUp. However, the designs generated by LC-FDS consumed more power (an average increase of 14.1%) than LegUp for segmentation. This increased power consumption was a tradeoff, as LC-FDS reduced the area by up to 77.6% for the non-unrolled segmentation. For feature extraction, LC-FDS reduced the power consumption and area by 11.6% and 71.5%, respectively, compared to LegUp.

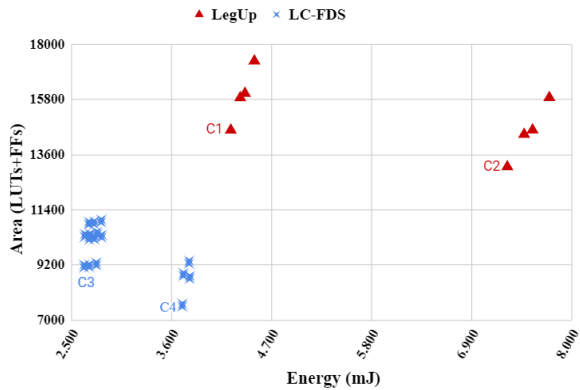

Fig. 8. Area (LUTs+FFs) vs. energy (mJ) design points for the wearable pregnancy monitoring device's MHR, SPO2, and EMG MCC hardware implementation combinations using the LegUp (triangular points) and LC-FDS (X points) HLS tools. C1, C2, C3, and C4 represent the Pareto-optimal design points for each tool.

### 5.2. Timing-constrained design space exploration

Given the Pareto-optimal hardware implementations of MCCs generated from the LC-FDS and LegUp tools and the timing specification within PSM models, we performed a timing-constrained, system-level design space exploration for the systems in our case studies. We explored the resulting energy and area tradeoffs for all combinations of MCC hardware implementations using a common operating frequency for all MCCs. For the WPM device, the design space exploration yielded eight possible system designs for LegUp and 32 possible designs for LC-FDS. For each combination, the common operating frequency was scaled down such that each MCC executed only as fast as needed based on the PSMs' periods. For instance, even though the frequency for SPO2 was 102 MHz, the frequency can be scaled down to just 49.5 KHz, which is fast enough for the MCC to run in one execution cycle of the SPO2 PSM.

Fig. 8 plots the resulting total area and energy consumptions of the hardware implementations for the WPM device. By reducing the frequency to execute the MCC only as fast as needed by the PSMs, the system-level energy consumption was reduced by 93.5% for LC-FDS and by 91.0% for LegUp, compared to the energy consumption when the frequency is unscaled.

LC-FDS yielded lower energy consumption and area than LegUp. Two critical points contribute to this observation. Firstly, LC-FDS was able to achieve much lower scaled frequencies than LegUp. For instance, the highest scaled frequency for the LC-FDS design points was 8.85 MHz, whereas the lowest scaled frequency for LegUp design points was 10.7 MHz. With lower scaled frequencies, the LC-FDS design points achieved lower energy consumption. Secondly, for each MCC hardware implementation, LC-FDS MCCs had lower area. As such, LC-FDS consumed less total area for the system-level implementation.

LC-FDS and LegUp exhibited a similar distribution of energy and area tradeoffs, with half of the design points within a narrow range. For example, half of the design points of LC-FDS were between 2.635 and 2.830 mJ and the other half were between 3.714 to 3.796 mJ. This distribution occurs because even though each configuration's frequency was scaled to run only as fast as needed, the EMG MCC requires a relatively long execution time. As a result, the scaled frequency was limited by the frequency required by the EMG MCC's. The average scaled frequency for the unrolled EMG MCC was 6.30 MHz, while the scaled frequency for the non-unrolled EMG MCC was 8.85 MHz.

To evaluate the overall H-HLS methodology, we compared the tradeoffs achieved by the Pareto-optimal configurations from the design space exploration using LC-FDS and LegUp. These Pareto-optimal points are denoted in Fig. 8 as *C1* and *C2* for LegUp and *C3* and *C4* for LC-FDS. The details for each of these Pareto-optimal configurations are as follows:

- *C1*: LegUp, MHR (unrolling: 0; maximum frequency: 121 MHz), SPO2 (unrolling: 0; Freq: 119 MHz), EMG (unrolling: 5; maximum frequency: 122 MHz).

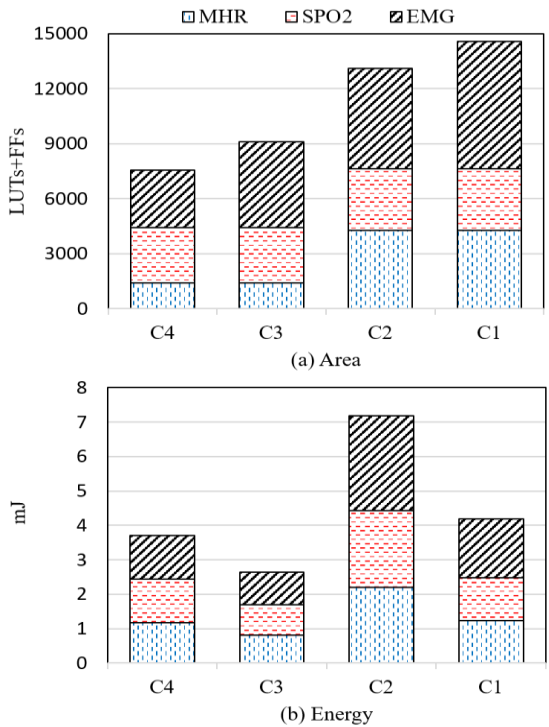

Fig. 9. Breakdown of (a) area (LUT+FF) and (b) energy (mJ) results of the Pareto-optimal system-level implementations for the WPM device using LegUp (C1 and C2) and LC-FDS (C3 and C4) HLS tools.

- *C2*: LegUp, MHR (unrolling: 0; maximum frequency: 121 MHz), SPO2 (unrolling: 0; maximum frequency: 119 MHz), EMG (unrolling: 0; maximum frequency: 120 MHz).

- *C3*: LC-FDS, MHR (unrolling: 0; latency constraint: 63; maximum frequency: 102 MHz), SPO2 (unrolling: 0; latency constraint: 56; maximum frequency: 103 MHz), EMG (unrolling: 5; latency constraint: 196; maximum frequency: 96 MHz).

- *C4*: LC-FDS, MHR (unrolling: 0; latency constraint: 63; maximum frequency: 102 MHz), SPO2 (unrolling: 0; latency constraint: 56; maximum frequency: 103 MHz), EMG (unrolling: 0; latency constraint: 163; maximum frequency: 94 MHz).

Among these Pareto-optimal configurations, C2 and C4 are the minimum area configurations for LegUp and LC-FDS, respectively. Conversely, C1 and C3 are the minimum energy configurations. C4 and C2 show that LC-FDS reduced the area and energy by 42.5% and 48.2%, respectively, compared to LegUp. For C3 and C1, LC-FDS reduced the area and energy by 37.6% and 38%, respectively. LC-FDS was able to use a lower frequency than LegUp to execute each MCC only as fast as needed given the execution periods defined for each PSM. By enabling a lower operating frequency via the precise specification of timing constraints, LC-FDS achieved greater energy savings compared to LegUp.

To derive further insights into how explicit timing specifications affect the system-level implementations, we analyzed the impacts of each MCC component on the overall area and energy consumption of the different Pareto-optimal configurations. Fig. 9 presents the breakdown of (a) area and (b) energy consumption of the four Pareto-optimal system-level implementations achieved by both LC-FDS and LegUp. For both PD-HLS tools, the area for MHR and SPO2 MCCs were the same between the minim-area and minimum energy Pareto-optimal configurations. The difference within the Pareto-optimal configurations occurred due to the impact of the EMG MCC. The configuration with smaller area used the non-unrolled EMG component, whereas the configuration with lower energy used the unrolled EMG component. For example, the non-unrolled EMG MCC in configuration C4 had 33.5% smaller area than the unrolled EMG MCC in C3, enabling C4 to reduce the total area by 17.1% compared to C3.

One can also observe that between the LC-FDS and LegUp tools, the SPO2 MCC has relatively small impact on the overall area reduction, as the hardware synthesized by each tool is similar in size. Comparing the area breakdown of C2 and C4, LC-FDS reduced the area by 67.3%, 9.2%, and 43.6% for the MHR, SPO2, and EMG MCCs, respectively. While the highest percentage reduction in area is achieved for the MHR component, the EMG component requires the largest area.

As shown in Fig. 9 (b), for the minimum-energy configurations C1 and C3, LC-FDS reduced the energy by 33.8%, 29.4%, and 45.0% for the MHR, SPO2, and EMG MCCs, respectively. The energy breakdown of C2 and C4, which are the minimum-area configurations, shows that LC-FDS reduced the energy by 46.9%, 43.4%, and 53.2% for the MHR,

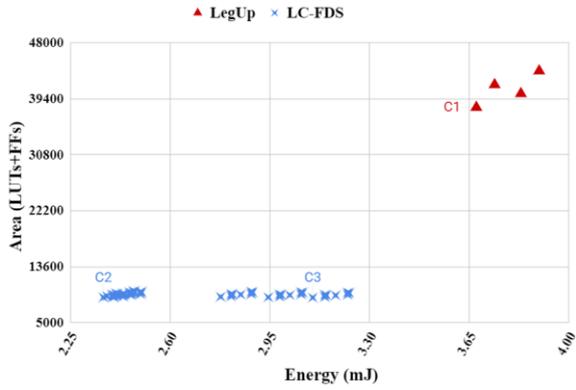

Fig. 10. Area (LUTs+FFs) vs. energy (mJ) design points for ECG-based biometric authentication system's filtering, segmentation, and feature extraction MCC hardware implementation combinations using the LegUp (triangular points) and LC-FDS (X points) HLS tools. C1, C2, and C3 represent the Pareto-optimal design points for each tool.

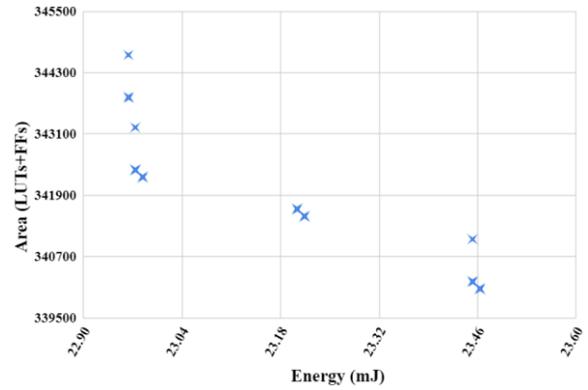

Fig. 11. Area (LUTs+FFs) vs. energy (mJ) Pareto-optimal design points for a synthetic system with four PSMs.

SPO2 and EMG PSMs. LC-FDS reduced the energy by the largest amount in the EMG component.

Even though C1 and C3 had the lowest energy consumption for both PD-HLS tools, they had larger areas compared to C2 and C4. For instance, C4 had the overall smallest area but had a concomitant energy increase of 29.1% compared to C3. C4's smaller area resulted from the non-unrolled EMG MCC, and the higher total energy consumption resulted from a higher scaled frequency to enable the system to run only as fast as needed for the PSMs. Conversely, the unrolled EMG component within C3 was selected based on the minimum scaled frequency achievable among the three alternative unrolled EMG MCC hardware implementations.

Our analysis of the EBA system and the synthetic system demonstrate similar benefits to the WPM device. For the EBA system, the design space exploration yielded four possible system combinations for LegUp and a total of 48 possible system combinations for LC-FDS. Fig. 10 plots the resulting total area and energy consumption of the hardware implementation combinations for the EBA device. Like the WPM device, the scaled frequency reduced the system-level energy consumption by 94.6% for LC-FDS and by 91.4% for LegUp, compared to the energy consumption of the system with an unscaled frequency. We identified the Pareto-optimal and optimal configurations for both LC-FDS and LegUp in Fig. 10. C2 and C3 are the Pareto-optimal points for LC-FDS, while C1 is the optimal point for LegUp. Given the small number of total configuration for LegUp, a single Pareto-optimal (and thus optimal) configuration was found.

The Pareto-optimal design points generated by LC-FDS (C2 and C3 in Fig. 10) achieved substantial area and energy reduction compared to LegUp's optimal design point (C1 in Fig. 10). Compared to LegUp's optimal point, C1, C2 reduced the area and energy by 75.6% and 35.6%, respectively. Similarly, C3 reduced the area and energy by 76.7% and 15.6%, respectively, compared to C1.

The generic synthetic system further illustrates the benefits of the H-HLS approach. As described in Table 2, the synthetic system had three, three, two, and three MCCs for each PSM. Each MCC has different number of MCC alternatives. Fig. 11 plots the resulting total area and energy consumption of the Pareto-optimal hardware implementation combinations for the synthetic system. We exhaustively searched all possible system configurations. The full design space exploration yielded more than 260 billion possible system configurations, of which there were 17 Pareto-optimal configurations, as shown in Fig. 11. Like the previously analyzed WPM and EBA systems, the system-level frequency scaling for the synthetic system reduced the system-level energy consumption by 90.5% compared to the unoptimized system. Overall, the explicit timing specification enabled by LC-FDS allowed different hardware implementations for the same MCCs leading to better tradeoff analysis and increased energy and area savings.

## 6. Conclusions and future work

High-level synthesis (HLS) has been widely used in designing embedded systems to enable increased productivity while reducing development costs. However, even though embedded systems typically have stringent and precise timing constraints, existing HLS approaches do not address the problem of precise timing specification during the synthesis and optimization process. In this paper, we introduced a hybrid high-level synthesis (H-HLS) methodology for synthesizing and optimizing application-specific embedded systems. H-HLS combines state-based HLS with performance-driven HLS and utilizes explicit and precise timing information within periodic state machine (PSM) models to effectively reduce energy consumption. We evaluated the benefits of the H-HLS methodology using case studies of the computations in a wearable pregnancy monitoring (WPM) device and an ECG-based biometric authentication (EBA) system. To further analyze the design space exploration in the proposed approach, we also analyzed a generic synthetic system with abstract computations. Our experiments demonstrate the ability of the H-HLS approach to generate hardware implementations that satisfy precise timing constraints while substantially reducing the system energy consumption. To quantify the benefits of the H-HLS methodology, we used two different PD-HLS tools, namely Microchip's LegUp HLS tool and a custom-built latency-constrained, force-directed scheduling (LC-FDS) HLS tool. Results reveal that, compared to an unoptimized design, the H-HLS methodology yields average energy reductions of 92.9%, and up to 94.6% for the EBA system. Furthermore, using LC-FDS, which supports explicit latency constraints and precise timing specifications within the system models, compared with LegUp which does not have this feature, the H-HLS approach reduces energy and area by an average of 41.9% and 59.6%, and up to 48.2% and 76.7%, respectively.

The work presented herein demonstrates the viability and benefits of the hybrid high-level synthesis approach to enable the design of precisely-timed energy-efficient embedded systems, but much future work remains. For larger and more complex systems, not only is automation required, but efficient design space exploration algorithms are needed. Future work includes exploring efficient and robust design space exploration methods, such as genetic algorithms or machine learning, to effectively explore the design space and obtain near-optimal solution for the timing-constrained DSE. We will also develop methods to automate the state-based HLS methodology within our LC-FDS HLS tool. For example, the system-level energy consumption and area estimate analysis and optimization can be automated within the tool. Beyond just supporting latency constraint specification for MCCs, the H-HLS approach can be extended to also support precisely-timed communications between PSMs. Additional future work includes exploring the benefits of the H-HLS approach in a wider variety of applications with more complex multi-cycle computations. Apart from healthcare wearables, several other fields can be explored, including, but not limited to, autonomous vehicles, agriculture, and traffic systems.